\documentclass[12pt,a4paper,final]{iopart}

\usepackage{iopams}  
\usepackage[breaklinks=true,colorlinks=true,linkcolor=blue,urlcolor=blue,citecolor=blue]{hyperref}
\usepackage[pdftex]{graphicx}
\usepackage{subfigure}
 \usepackage{tikz}
\usetikzlibrary{decorations.markings}

\def\beq{\begin{equation}}
\def\eeq{\end{equation}}
\def\bsp{\begin{split}}
\def\esp{\end{split}}
\def\bea{\begin{eqnarray}}
\def\eea{\end{eqnarray}}
\def\ba{\begin{array}}
\def\ea{\end{array}}

\def\dg{\dagger}

\def\lb{\left(}
\def\rb{\right)}

\def\l.{\left.}
\def\r.{\right.}

\def\ra{\rangle}
\def\la{\langle}

\def\bo{\bold{k}}

\begin{document}

\title[ Honeycomb topological magnon insulator ]{ A first theoretical realization of honeycomb topological magnon insulator}

\author{S. A. Owerre $^{1,2}$}

\address{$^1$Perimeter Institute for Theoretical Physics, 31 Caroline St. N., Waterloo, Ontario N2L 2Y5, Canada.}
\address{$^2$ African Institute for Mathematical Sciences, 6 Melrose Road, Muizenberg, Cape Town 7945, South Africa.}
\ead{solomon@aims.ac.za;~sowerre@perimeterinstitute.ca}

%

\begin{abstract}
   It has been recently  shown that  in the Heisenberg (anti)ferromagnet on the honeycomb lattice,   the magnons (spin wave quasipacticles) realize a massless two-dimensional (2D) Dirac-like Hamiltonian.  It was shown that the Dirac magnon Hamiltonian  preserves time-reversal symmetry defined with the sublattice pseudo spins and  the Dirac points are robust against magnon-magnon interactions. The Dirac points also  occur  at nonzero energy. In this paper, we propose a simple realization of  nontrivial topology (magnon edge states)  in this system. We show that the Dirac points are gapped when the inversion symmetry of the lattice is broken by introducing a next-nearest neighbour Dzyaloshinskii-Moriya (DM) interaction. Thus, the system  realizes  magnon edge states similar to Haldane model for quantum anomalous Hall effect in  electronic systems. However, in contrast to electronic spin current where dissipation can be very large due to Ohmic heating, noninteracting topological magnons can propagate for long time without dissipation as  magnons are uncharged particles.  We observe the same  magnon edge states for the XY model on the honeycomb lattice. Remarkably, in this case the model maps to interacting hardcore bosons on the honeycomb lattice. Quantum magnetic systems with nontrivial magnon edge states are called topological magnon insulators. They have been studied theoretically on the kagome lattice and recently observed experimentally  on the kagome magnet  Cu(1-3, bdc) with three magnon bulk bands. Our results for the honeycomb lattice suggests an experimental procedure to search for honeycomb topological magnon insulators within a class of 2D quantum magnets and ultracold atoms trapped in honeycomb optical lattices. In 3D lattices, Dirac and Weyl points  were recently studied theoretically, however, the criteria that give rise to them were not well-understood. We argue that  the low-energy Hamiltonian near the Weyl points should break time-reversal symmetry of the pseudo spins. Thus, recovers the same criteria in electronic systems.
\end{abstract}

\vspace{2pc}

\submitto{\ J. Phys. Condens. Matter}

\section{Introduction}

Topological properties of fermion band theory have dominated research in condensed matter physics and other areas over the past decade  or so \cite{has1, has2, cas,wang1, aab1, Xu}. Recently, it has been  shown that  the magnon bulk bands of Heisenberg (anti)ferromagnet on the honeycomb lattice exhibit Dirac points at the corners of the Brillouin zone (BZ) \cite{jf}.  The low-energy Hamiltonian near these points realizes a massless 2D Dirac-like Hamiltonian with Dirac nodes at nonzero energy. This system   preserves pseudo spin time-reversal ($\mathcal{T}$) symmetry. It was also shown that the Dirac points are robust against magnon-magnon interactions and   any perturbation that preserves the pseudo spin $\mathcal{T}$-symmetry of the Bogoluibov Hamiltonian. 
 
 In this paper, we provide evidence of non-trivial topology (magnon edge states) in the magnon bulk bands of   Heisenberg (anti)ferromagnet  and XY  model on the honeycomb lattice, when a gap opens at the Dirac points.  We show that the simplest practical way to open a gap at the Dirac points is by breaking the inversion symmetry of the lattice, which subsequently breaks the pseudo spin $\mathcal{T}$-symmetry of the Bogoliubov Hamiltonian. We show that this can be achieved by introducing  a next-nearest neighbour Dzyaloshinskii-Moriya (DM) interaction. The opening of a gap at the Dirac points leads to magnon edge states  reminiscent of  Haldane model in electronic systems \cite{adm}.    In the case of XY model, we observe the same topological  effects with magnon edge states propagating in the vicinity of the magnon bulk gap. Remarkably,  the resulting Hamiltonian for the  XY model maps to interacting hardcore bosons. Therefore, these magnon edge states can be simulated numerically.   As magnons are uncharged particles, noninteracting topological magnons can propagate for a long time without dissipation, thus they are considered as a good candidate for magnon spintronics \cite{kru,kru1,kru2,shin,shin1}.   The topological properties of these Dirac magnons are not just analogues of fermion band theory. They are called topological magnon insulators \cite{zhh, zhh1} and has been recently observed on the kagome magnet  Cu(1-3, bdc)\cite{alex6a}.  For the honeycomb lattice, they can actually be searched for  in many accessible quantum magnets such as Na$_3$Cu$_2$SbO$_6$ \cite{aat1} and $\beta$-Cu$_2$V$_2$O$_7$ \cite{aat} which are spin-$1/2$ Heisenberg antiferromagnetic materials  with a honeycomb structure. They can also be investigated in ultra-cold atoms trapped in honeycomb optical lattices as the bosonic tight binding model is analogous to that of  Haldane model,  which  has been realized experimentally in optical fermionic lattice \cite{jott}. 
  
  In 3D quantum magnets, Dirac and  Weyl points  are possible in the magnon excitations. Recently, Weyl points have been investigated in  quantum magnets using 3D Kitaev fermionic model  \cite{aab0}. Weyl points were recently shown to occur in the magnon excitations of  breathing pyrochlore lattice antiferromagnet \cite{fei, up}.  In this case, the criteria that give rise to them seem to be unknown unlike in fermionic systems.   Here, we show that the resulting Bogoliubov Hamiltonian has the form of Weyl Hamiltonian in electronic systems and that the  Weyl points should break the pseudo spin $\mathcal{T}$-symmetry by expanding the Bogoliubov Hamiltonian near the non-degenerate dispersive band-touching points and projecting onto the bands. Hence, the criterion for Weyl nodes to exist in electronic systems also applies to magnons.
  
\section{Honeycomb Dirac Magnon}  
  
\begin{figure}[ht]
\centering
\includegraphics[width=4in]{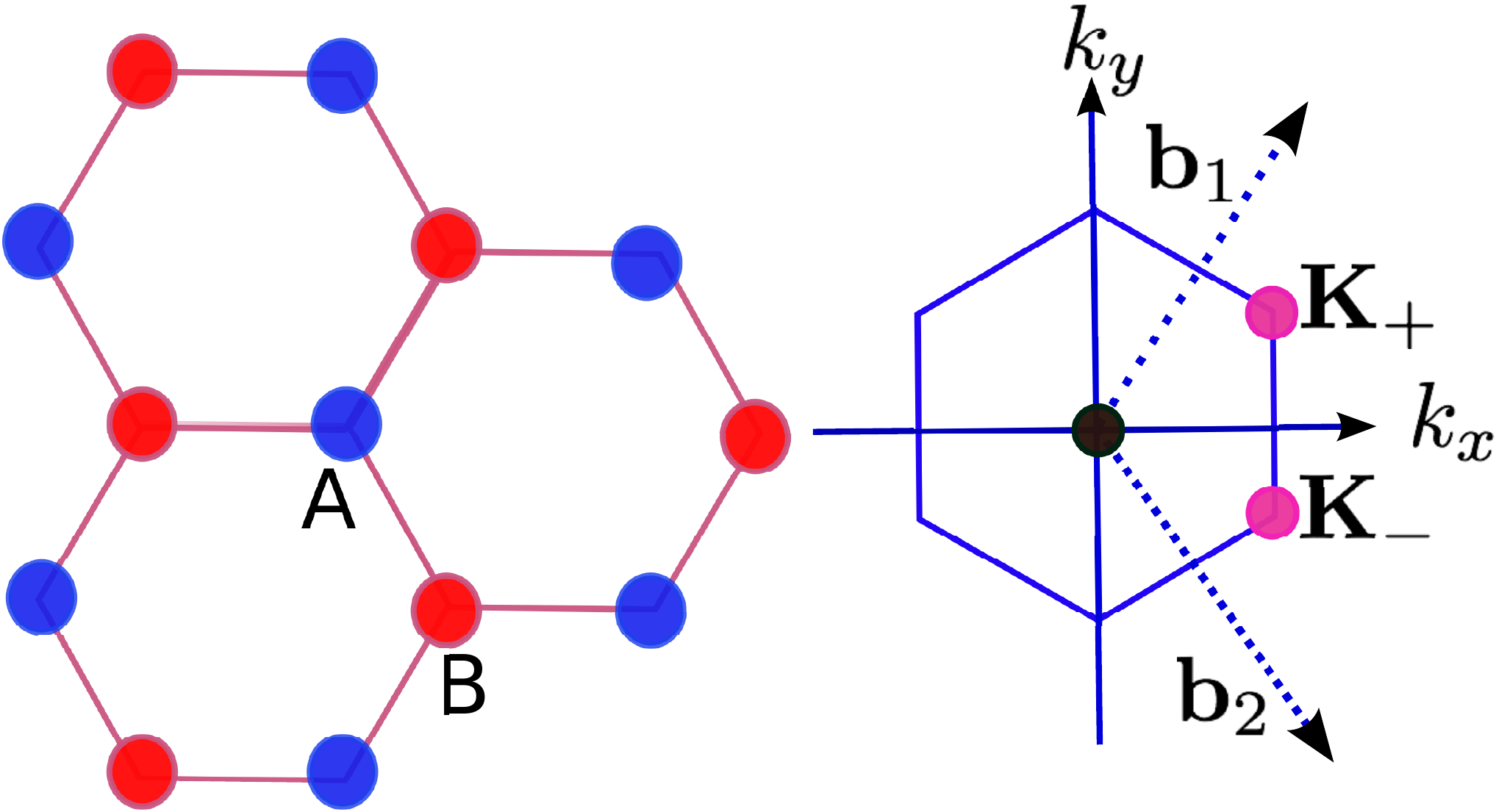}
\caption{Color online.  The  honeycomb lattice (left) and the Brillouin zone (right).  The reciprocal lattice vectors are ${\bf b}_1= 2\pi(1,\sqrt{3})/3a$ and ${\bf b}_2= 2\pi(1,-\sqrt{3})/3a$.}
\label{honey_comb}
\end{figure}

In 2D quantum spin magnetic materials, non-degenerate band-touching points or Dirac points  require at least two energy branches of the magnon excitations. Therefore, ordered quantum magnets that can be treated with one sublattice  are devoid of Dirac points. The simplest two-band model that exhibits  Dirac nodes is the Heisenberg ferromagnet or antiferromagnet on the honeycomb lattice \cite{jf}.  The Hamiltonian is governed by
\begin{eqnarray}
&H=-\sum_{\la lm\ra}J_{lm}{\bf S}_{l}\cdot{\bf S}_{m},
\label{hhh1}
\end{eqnarray}
where $J_{lm}$ depends on the bonds along the nearest neighbours.   As mentioned above, Eq.~\ref{hhh1} describes several realistic compounds \cite{aat1, aat}. For simplicity we take $J_{lm}=J>0$. The ground state  of Eq.~\ref{hhh1} is a ferromagnet with  two-sublattice structure on the honeycomb lattice; see Fig.~\ref{honey_comb}. This is equivalent to Heisenberg antiferromagnet by flipping the spins on one sublattice.  In many cases of physical interest,  magnon excitations are studied by linear spin wave theory via the standard linearize Holstein Primakoff (HP) transformation. This is an approximation valid at low-temperature  when few magnons are excited. 
\begin{figure}
\centering
  \subfigure[\label{HCA}]{\includegraphics[width=.45\linewidth]{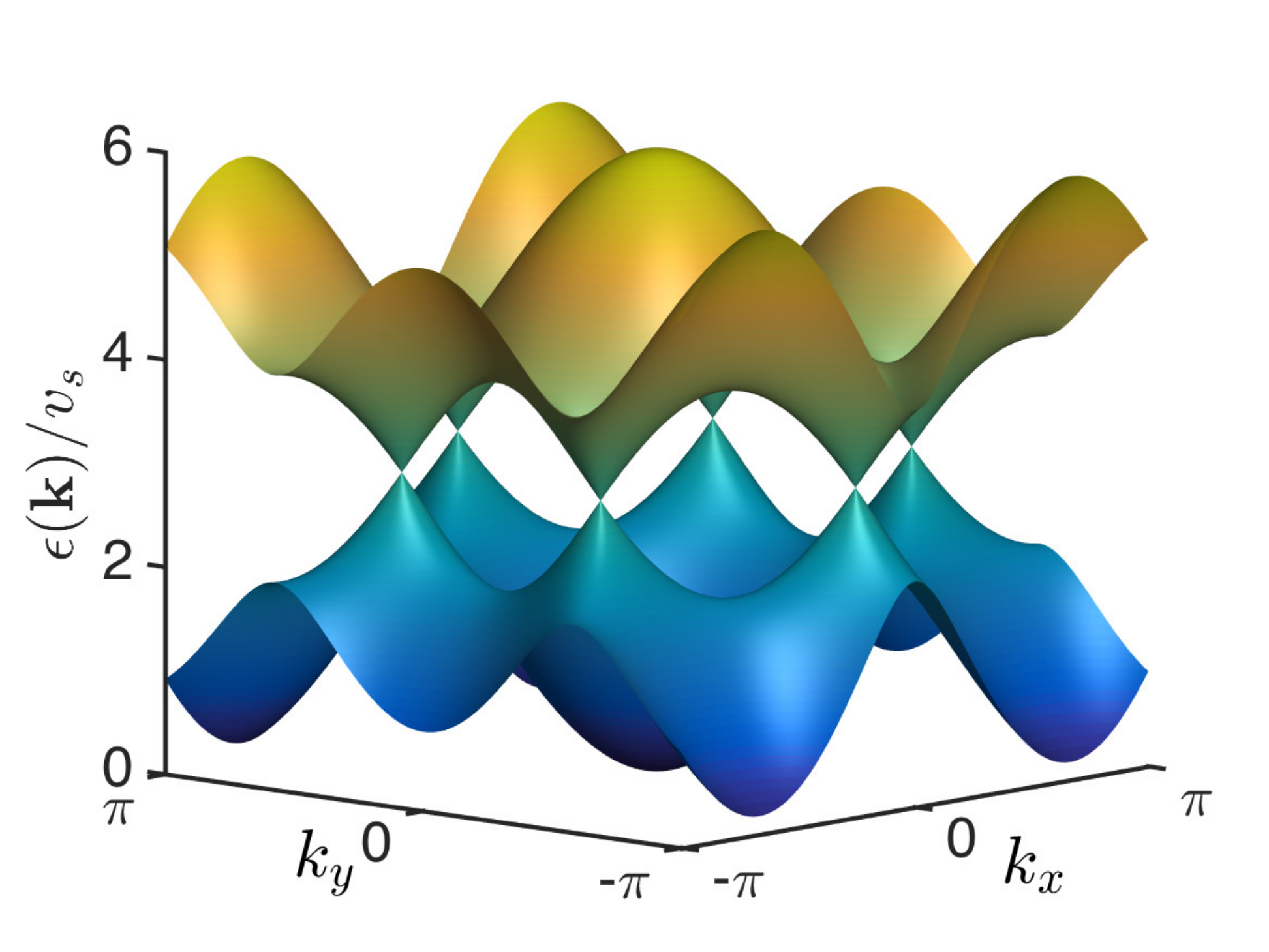}}
   \quad
   \subfigure[\label{DOS}]{\includegraphics[width=.45\linewidth]{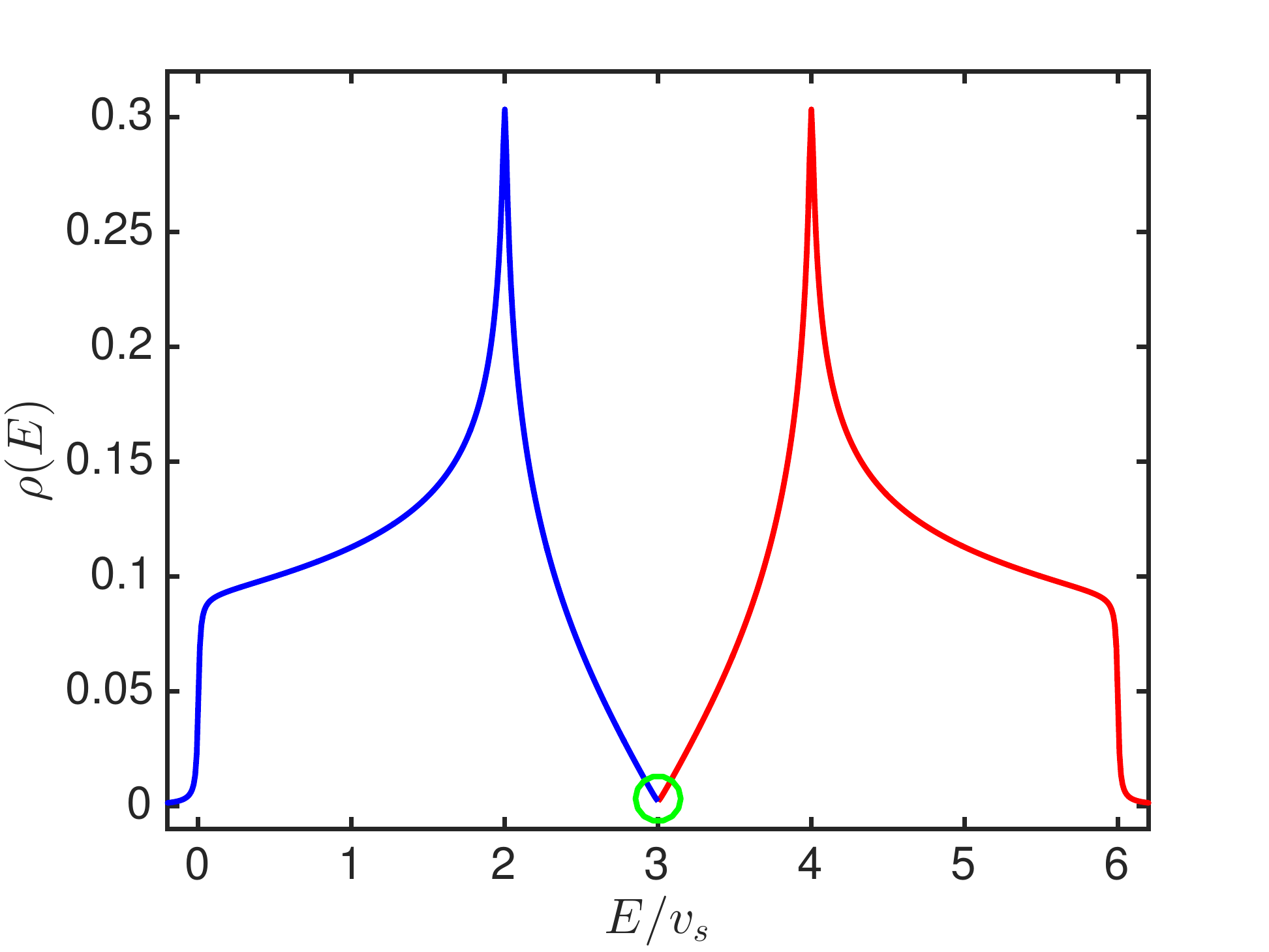}}
\caption{Color online. $(a)$ The  energy bands  of the Heisenberg honeycomb ferromagnet.  $(b)$ Density of states per unit cell  of the Heisenberg ferromagnet on the honeycomb lattice. The blue and red lines denote the two bands and the green circle is the point of degeneracy with $E=3v_s$.}
\end{figure}

%
%

 The  momentum space Hamiltonian is given by $H=\sum_{\bo}\Psi^\dg_\bo\cdot \mathcal{H}_B(\bo)\cdot\Psi_\bo $, where $\Psi^\dg_\bo= (a_{\bo}^{\dg},\thinspace b_{\bo}^{\dg})$ and the mean field energy has been dropped. In this model, the Bogoliubov quasiparticle operators are the same as the bosonic operators.   The Bogoliubov Hamiltonian is given by  
\begin{eqnarray}
\mathcal{H}_B(\bo)&=zv_s\sigma_0- zv_s(\sigma_+\gamma_\bo +h.c.),
\label{honn}
\end{eqnarray}
where  $\sigma_0$ is an identity $2\times 2$ matrix,  and  $\sigma_\pm=(\sigma_x\pm i\sigma_y)/2$ are Pauli matrices acting on the sublattices;   $z=3$ is the coordination number of the lattice and  $v_s=JS$. 
 The structure factor $ \gamma_\bo$ is complex given by
\begin{eqnarray}\gamma_\bo=\frac{1}{z}\sum_{\mu} e^{i\bo\cdot \boldsymbol{\delta}_\mu},\end{eqnarray}
where  $ \boldsymbol{\delta}_\mu$ are the three nearest neighbour vectors on the honeycomb lattice, $ \boldsymbol{\delta}_1=(\hat x,\sqrt{3}\hat y)/2$, $ \boldsymbol{\delta}_2=(\hat x,-\sqrt{3}\hat y)/2$ and $ \boldsymbol{\delta}_3=(-\hat x, 0)$.
The eigenvalues of Eq.~\ref{honn} are given by
\begin{eqnarray}
\epsilon_\pm({\bo})= 3{v}_s\lb1 \pm |\gamma_\bo|\rb.
\label{sfk}
\end{eqnarray}
The energy bands have Dirac nodes at the corners of the BZ  reminiscent of graphene model. In contrast to graphene,  the Dirac nodes  occur with a nonzero energy $3{v}_s$ as shown in Fig.~\ref{HCA}. In addition to the Dirac nodes, there is a zero energy mode in the lower band,  which corresponds to a Goldstone mode due to the spontaneous symmetry breaking of SU(2) rotational symmetry of the quantum spin Hamiltonian. As many physical systems are anisotropic, it is important to note that with spatial anisotropy $J_{lm}\neq J$, several Dirac points can be obtained by tuning the anisotropy in each bond. The density of states per unit cell as a function of energy is shown in Fig. \ref{DOS} for the two energy bands in Eq.~\ref{sfk}. The interesting properties of this system are manifested near the Dirac points.  There are only two inequivalent Dirac points located at $\bold{K}_\pm=\lb{2\pi}/{3}, \pm{2\pi}/{3\sqrt{3}}\rb$ as shown in Fig.~\ref{honey_comb}. In the case of Heisenberg antiferromagnet, only a single Dirac point occurs at $\bo=0$ \cite{jf}.   Expanding  Eq.~\ref{honn} near $\bold{K}_\pm$ we obtain a linearized model 

 \begin{eqnarray}
\mathcal{H}_B(\bold q)&=3{v}_s\sigma_0+ \tilde{v}_s(\sigma_x q_x- \tau\sigma_y q_y),
\label{me1}
\end{eqnarray}
where $\bold{q}=\bo-\bold{K}_\pm$, $\tilde{v}_s={v}_s/2$, and $\tau=\pm$ describes states at $\bold{K}_\pm$. Thus, the low-energy  excitation spectrum near the Dirac points is similar to the Bloch Hamiltonian of graphene model. Let us now compute the specific heat at a constant volume given by
\bea
c_v=\lb\frac{\partial u}{\partial T}\rb_V,
\eea
where $u$  is the internal energy density of the system given by
\bea
u=\sum_{\lambda=\pm}\int \frac{d^2q}{(2\pi)^2} \epsilon_\lambda(\bold q)n_B[\xi_\lambda(\bold q)],
\eea
$\xi_\lambda(\bold q)=\epsilon_\lambda(\bold q)-\mu$, $n_B[\epsilon_\lambda(\bold q)]=[e^{\beta \xi_\lambda(\bold q)}-1]^{-1}$ is the Bose function, $\beta=1/T$ is the inverse temperature, and $\mu$ is the chemical potential. The specific heat can be integrated exactly by turning the chemical potential at the Dirac nodes $\mu=3v_s$. In this case we find
\begin{eqnarray}
c_v= \frac{v_s^2}{2T^2}\int \frac{d^2q}{(2\pi)^2}\frac{|\gamma(\bold q)|^2}{\sinh^2\lb\frac{v_s|\gamma(\bold q)|}{2T}\rb}=\frac{12\zeta(3)}{\pi v_s^2}T^2,
\end{eqnarray}
where $\zeta(3)=1.20206$. This is similar to the $T^2$-law found in graphene.

A very crucial point is the role of time-reversal symmetry.  Since the  excitations of quantum magnets are usually described in terms  of  the HP bosons, an ordered state must be assumed.  Hence,  the system must contain an even number of  half integral spins  with $\mathcal T^2=(-1)^N$, where $N$ is even.   Thus, magnons behave like bosons. However, in the pseudo spin space  ${\mathcal T}$-operator can be defined for the Bogoliubov Hamiltonian,  ${\mathcal T}=i\sigma_y\mathcal K$ where $\mathcal K$ denotes complex conjugation and ${\mathcal T}^2=-1$.  This pseudo spin symmetry is preserved  provided Dirac points exist in the BZ. 

\section{Honeycomb Topological Magnon Insulator}

\subsection{Heisenberg ferromagnetic insulator}
Topological magnon insulators are the analogues of topological insulators in electronic systems.  They are characterized by the existence of edge state modes when a gap opens at the Dirac points.   For the honeycomb ferromagnets, a  next-nearest neighbour interaction of the form $H=-J^\prime\sum_{\la \la lm\ra\ra}{\bf S}_{l}\cdot{\bf S}_{m}$ ($J^\prime>0$) only shifts the positions of the Dirac points as it contributes a term of the form $v_s^\prime(6-g_\bo)\sigma_0$, where  $v_s^\prime=J^\prime S$, and $g_\bo=2\sum_\mu\cos \bo\cdot\boldsymbol{\rho}_\mu$.   The next-nearest neighbour vectors are $ \boldsymbol{\rho}_1=-(3\hat x,\sqrt{3}\hat y)/2$, $ \boldsymbol{\rho}_2=(3\hat x,-\sqrt{3}\hat y)/2$, $ \boldsymbol{\rho}_3=(0,\sqrt{3}\hat y)$. The  simplest  realistic way to open a gap at the Dirac points is by breaking the inversion symmetry of the lattice, which in turn breaks the ${\mathcal T}$-symmetry of the Bogoliubov Hamiltonian. This can be achieved by introducing a next-nearest neighbour  DM interaction
\begin{eqnarray}
H_{DM}= \sum_{\la \la lm\ra\ra}{\bf D}_{lm}\cdot{\bf S}_{l}\times{\bf S}_{m},
\label{h1}
\end{eqnarray}
where ${\bf D}_{lm}$ is the DM interaction between sites $l$ and $m$.  The total Hamiltonian of a honeycomb ferromagnetic insulator can be written as
\begin{eqnarray}
H= -J\sum_{\la lm\ra}{\bf S}_{l}\cdot{\bf S}_{m}-J^\prime\sum_{\la \la lm\ra\ra}{\bf S}_{l}\cdot{\bf S}_{m}+\sum_{\la \la lm\ra\ra}{\bf D}_{lm}\cdot{\bf S}_{l}\times{\bf S}_{m}.
\end{eqnarray}

\begin{figure}[ht]
\centering
\includegraphics[width=2in]{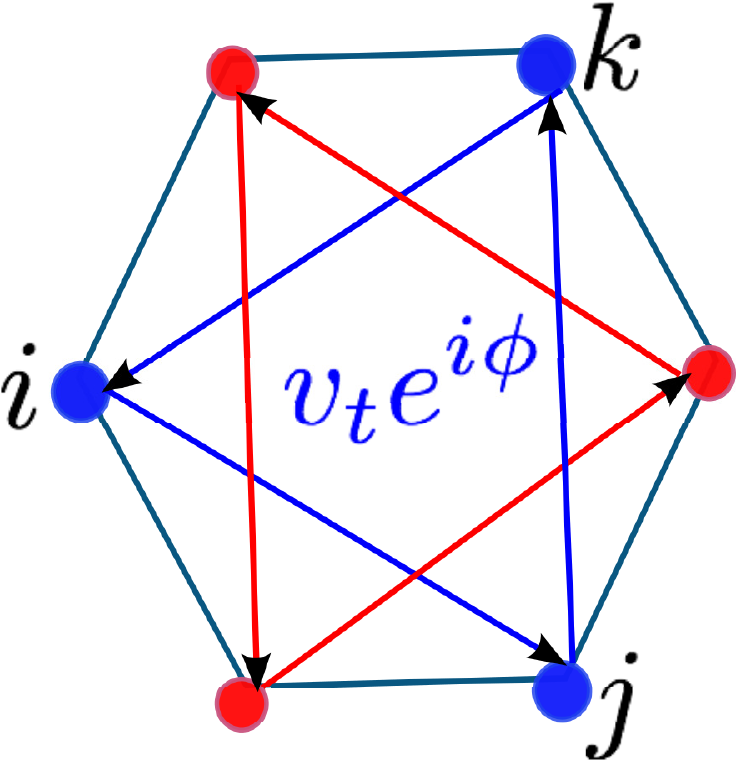}
\caption{Color online.  The magnetic flux treads on the honeycomb lattice  generated by a NNN DM interaction;  $i,~j,~k$ label sites on the triangular plaquettes and give rise to a nonzero spin chirality $\chi={\bf S}_i\cdot {\bf S}_j\times {\bf S}_k$.}
\label{unit_cell_1}
\end{figure}

In the HP bosonic mapping, we obtain
\begin{eqnarray}
H&=-v_s\sum_{\la lm\ra}( b_l^\dagger b_m+ h.c.)  - v_t\sum_{\la \la lm\ra\ra}(e^{i\phi_{lm}} b^\dagger_l b_{m}+h.c.) +v_0\sum_l b_l^\dagger b_l,
\label{hp3}
\end{eqnarray}
where  $v_0=zv_s+z^\prime v_s^\prime$, $v_D=DS$, $v_t=\sqrt{v_s^{\prime 2} +v_D^2}$, and $z^\prime=6$ is the coordination number of the NNN sites. We have assumed a DM interaction along the $z$-axis.   The phase factor $\phi_{lm}=\nu_{lm}\phi$, where $\phi=\arctan(D/J^\prime)$ is a magnetic flux generated by the DM interaction on the NNN sites, similar to the Haldane model with $\nu_{lm}=\pm 1$ as in electronic systems.   The total flux enclosed in a unit cell vanishes as depicted in Fig.~\ref{unit_cell_1}. In contrast to electronic systems, the phase factor $\phi$ depends on the parameters of the Hamiltonian.
The Bogoliubov Hamiltonian is given by  
\begin{eqnarray}
\mathcal{H}_B(\bold k)&=h_0\sigma_0 +h_x\sigma_x + h_y\sigma_y + h_z\sigma_z,
\end{eqnarray}
where $h_0= 3v_s-2v_t\cos\phi g_\bo$, $h_x=-v_s\sum_\mu\cos \bo\cdot\boldsymbol{\delta}_\mu$, $h_y=-v_s\sum_\mu\sin \bo\cdot\boldsymbol{\delta}_\mu$, and $h_z=-2v_t\sin\phi \rho_\bo$, where $\rho_\bo=\sum_\mu\sin \bo\cdot\boldsymbol{\rho}_\mu$ .
Expanding near the Dirac points we  generate a gap at ${\bf K}_\pm$ and the full Hamiltonian for $J^\prime =0$ ($\phi=\pi/2$ in this case) becomes
 \begin{eqnarray}
\mathcal{H}_B(\bold q)&=3{v}_s\sigma_0+ \tilde{v}_s(\sigma_x q_x- \tau\sigma_y q_y) +m\tau\sigma_z,
\label{mme1}
\end{eqnarray}
where $m=3\sqrt{3}v_D$. 
\begin{figure}
\centering
  \subfigure[\label{Edge}]{\includegraphics[width=.45\linewidth]{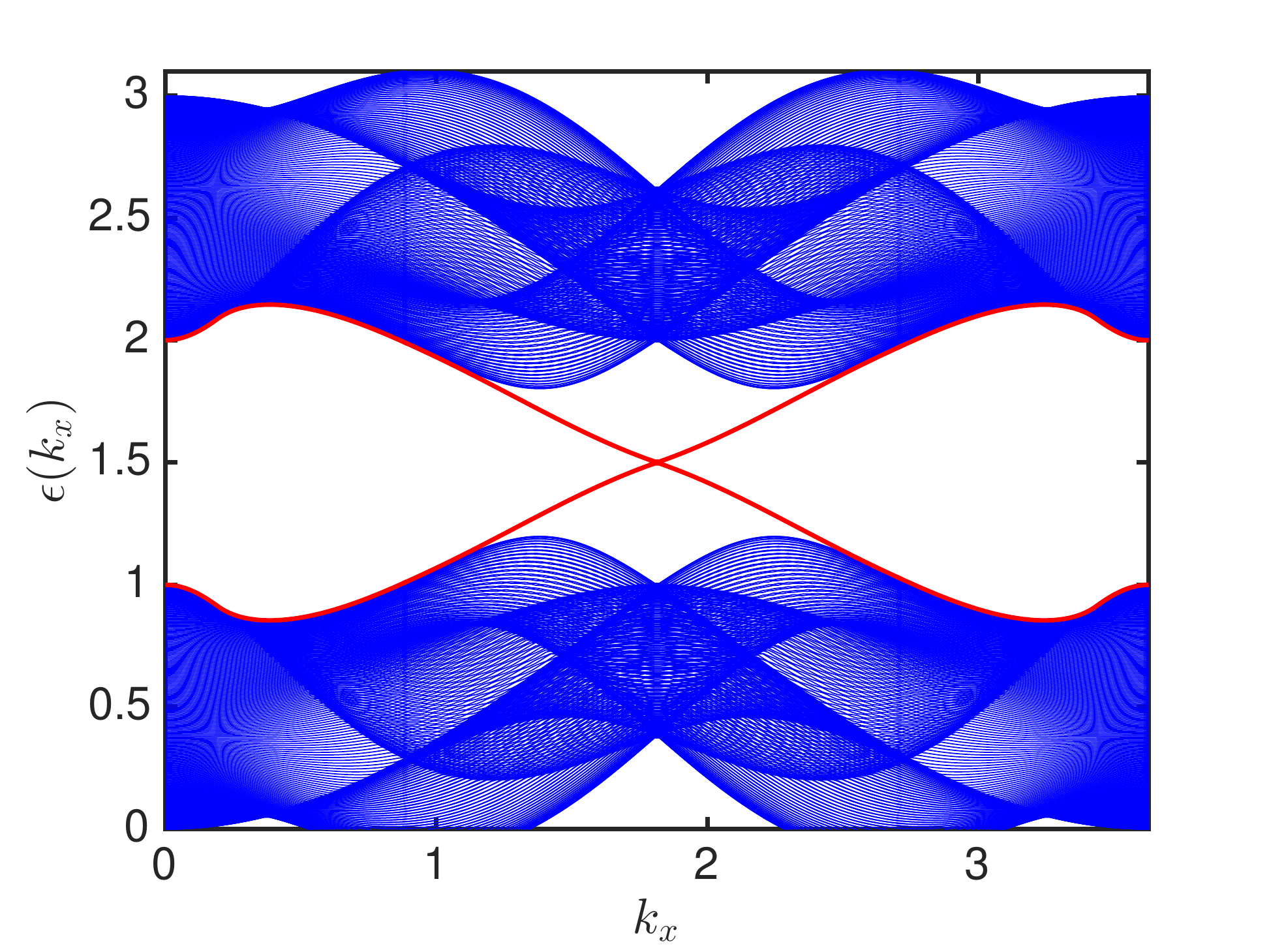}}
   \quad
   \subfigure[\label{Edge1}]{\includegraphics[width=.45\linewidth]{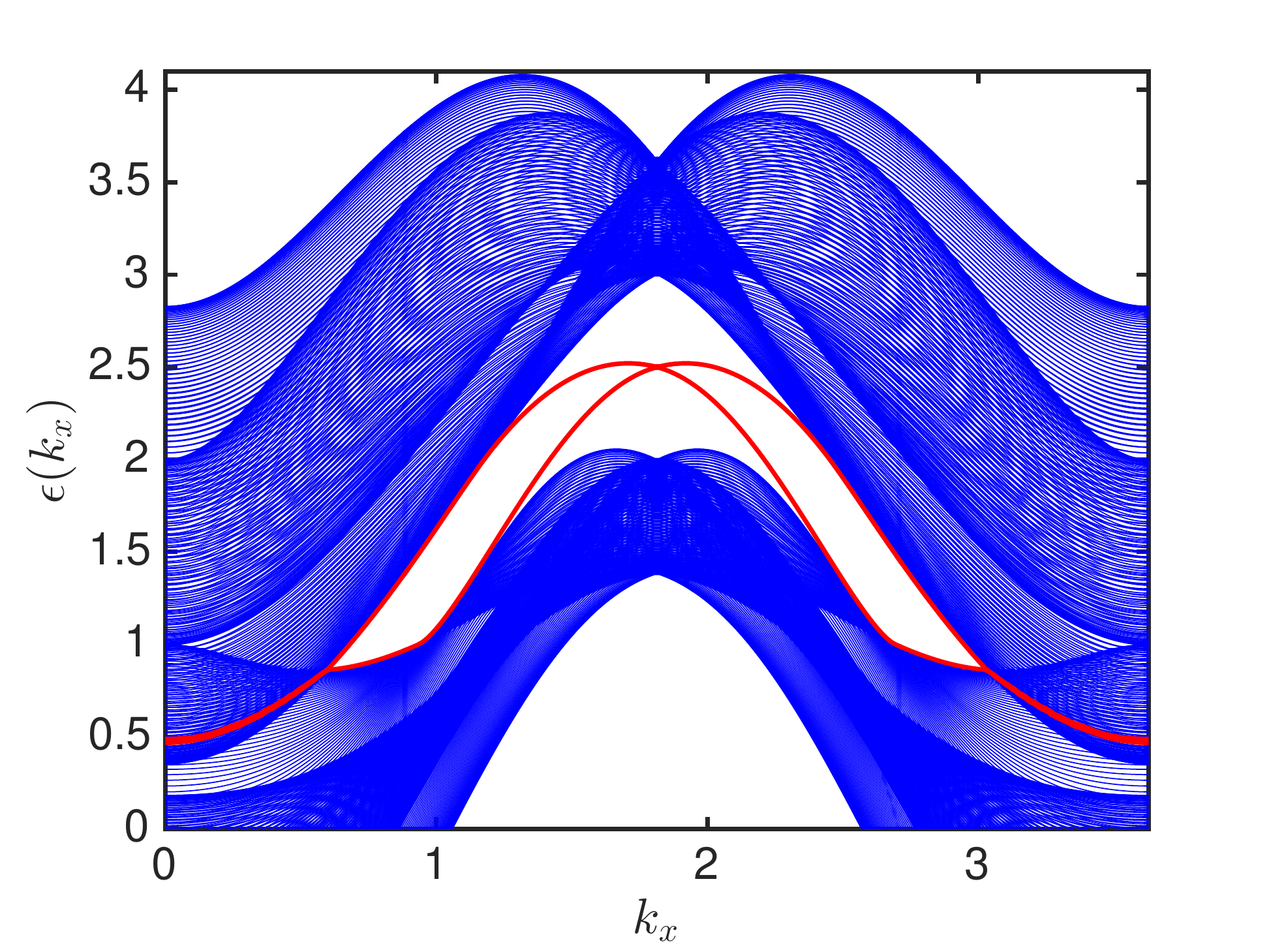}}
\caption{Color online. The energy band for a one-dimensional strip on the honeycomb lattice for spin-$1/2$ in units of $J=1$. $(a)$ $J^\prime=0, D=0.5J,~\phi=\pi/2$. $(b)$ $J^\prime=D=0.5J,~\phi=\pi/4$. }
\end{figure}
\begin{figure}[ht]
\centering
\includegraphics[width=3in]{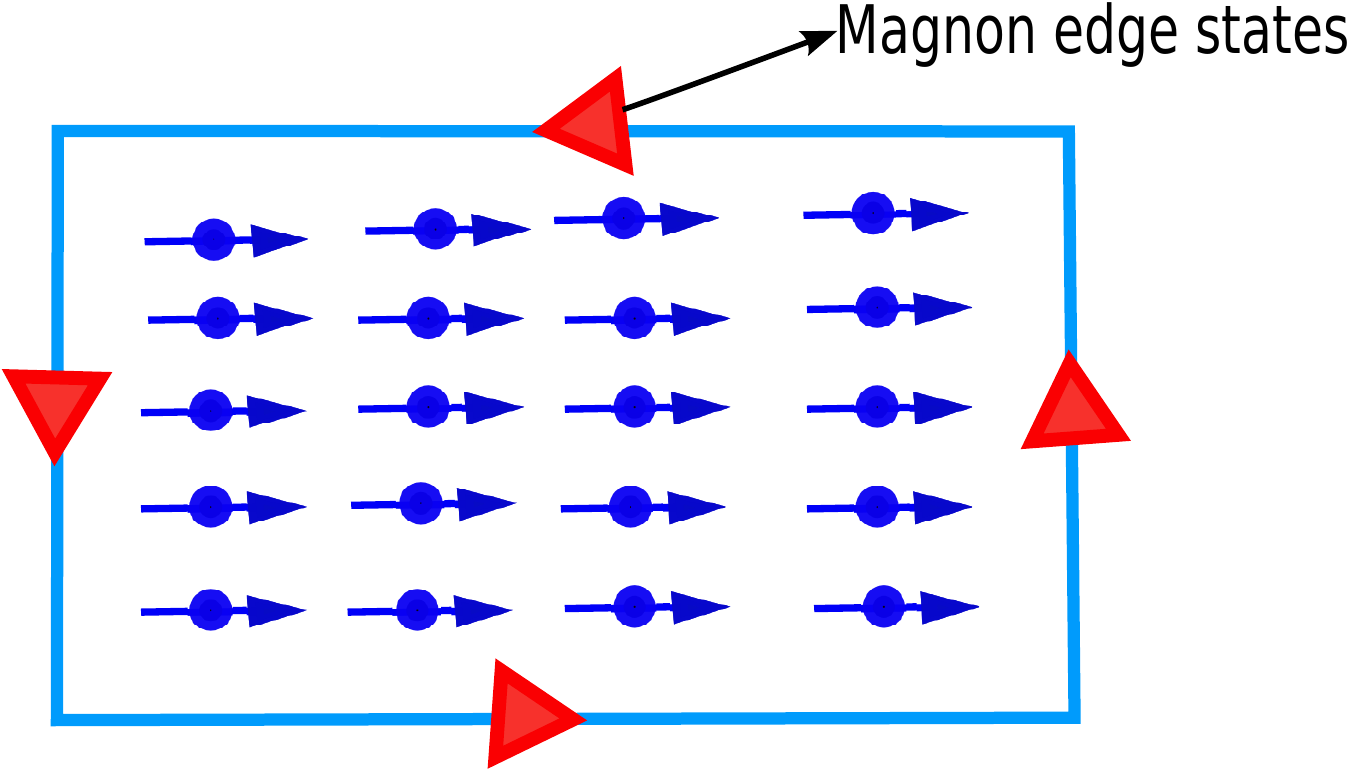}
\caption{Color online. Schematics of magnon edge states in topological magnon insulator material.}
\label{Edge2}
\end{figure}
This model can be regarded as  the bosonic analogue of Haldane model in electronic systems \cite{adm}. In  fermionic systems, there is a topological invariant quantity which is quantized when the Fermi energy lies between the gap, such that the lower band is occupied. In the bosonic model, there is no Fermi energy and not all states are occupied. The bosons can condense at the Goldstone mode in the lower band.  However, the topological invariant quantity  is, in principle,  independent of the statistical property of the particles. It merely predicts edge states in the vicinity of the bulk gap. In the magnon excitations the Chern number $n_H= \rm{sign}(m)$ simply predicts  a pair of counter-propagating magnon edge states in the vicinity of the bulk gap as shown in Figs.~\ref{Edge} and \ref{Edge1} for $\phi=\pi/2$ and $\phi=\pi/4$ respectively.   Hence, the Heisenberg (anti)ferromagnet on the honeycomb lattice realizes topological magnon insulator with magnon edge states propagating at the edge of the sample as depicted in Fig.~\ref{Edge2}.   As mentioned above, the propagation of magnon edge states differs from those in electronic systems.  Therefore, they a useful in many technological devices and magnon spintronics.  Besides, they can be accessible in many accessible quantum magnetic systems.

\subsection{XY ferromagnetic insulator: hardcore bosons}

Dirac points occur in a variety of quantum honeycomb ferromagnetic insulators. Let us consider  the XY model on the honeycomb lattice
\begin{eqnarray}
H=-2J\sum_{\langle lm\rangle}(S_l^xS_m^x+S_l^yS_m^y).
\label{ham0}
\end{eqnarray}
The ground state of this model is an ordered ferromagnet or N\'eel state in the $xy$ plane. Choosing $S_x$ quantization axis, the momentum space Hamiltonian in linear spin wave theory  is generally written as

\begin{eqnarray}
H=&\mathcal{E}_c+ \sum_{\bo,\mu,\nu}\mathcal{A}_{\mu\nu}(\bo) b_{\bo \mu}^\dagger b_{\bo \nu}\label{main} +\mathcal{B}_{\mu\nu}(\bo)  b_{\bo \mu}^\dagger b_{-\bo \nu}^\dagger+ \mathcal{B}_{\mu\nu}^*(\bo)b_{-\bo \mu} b_{\bo \nu},
\end{eqnarray}
where $\mu,\nu$ label the sublattices. Equation~\ref{main} can be written as
\bea H=\mathcal{E}_0+ \sum_{\bo}\Psi^\dg_\bo \cdot \mathcal{H}(\bo)\cdot\Psi_\bo +\rm{const.},
\label{hp}\eea where $\Psi^\dg_\bo= (\psi^\dg_\bo, \thinspace \psi_{-\bo} )$, $\psi^\dg_\bo=(b_{\bo 1}^{\dg}\thinspace b_{\bo 2}^{\dg}\cdots \thinspace b_{\bo N}^{\dg})$, and $N$ is the number of sublattice, $\mathcal{E}_0={\mathcal{E}_c}-S\sum_{\bo\mu}\mathcal{A}_{\mu\mu}(\bo)$, and  \bea \mathcal{H}(\bo) = \sigma_0 \otimes \boldsymbol{\mathcal{A}(\bo)}  + \sigma_+\otimes\boldsymbol{\mathcal{B}(\bo)}+ \sigma_- \otimes\boldsymbol{\mathcal{B^*}(\bo)},\eea where $\boldsymbol{\mathcal{A}}(\bo)$ and $\boldsymbol{\mathcal{B}}(\bo)$ are $N\times N$ matrices.  This Hamiltonian is diagonalized by a matrix $\mathcal{U}(\bo)$ via the transformation $\Psi^\dg_\bo\to\mathcal{U}(\bo)\mathcal{P}(\bo)$, which satisfies the relation \bea\mathcal{U}^\dg \mathcal{H}(\bo) \mathcal{U}= \epsilon(\bo); \quad \mathcal{U}^\dg \eta \mathcal{U}= \eta,\eea with $\eta=\rm{diag}(\mathbf{I}_{N\times N}, -\mathbf{I}_{N\times N} )$.  $\mathcal P(\bo)$ contains the Bogoliubov operators $\alpha_{\bo}^{\dg}$ and $ \beta_{\bo}^{\dg}$ and  $\epsilon(\bo)$ is the  eigenvalues. This is equivalent to saying that we need to diagonalize a non-hermitian Bogoliubov Hamiltonian  $\mathcal{H}_B(\bo)=\eta\mathcal{H}(\bo)$, where

 \begin{eqnarray}
\mathcal{H}_B(\bo)&=  \sigma_x\otimes\boldsymbol{\mathcal{B}}_-(\bo) +i \sigma_y\otimes\boldsymbol{\mathcal{B}}_+(\bo)+\sigma_z\otimes\boldsymbol{\mathcal{A}(\bo)},
\label{Bogoliubovb}
\end{eqnarray}
and $\boldsymbol{\mathcal{B}}_\pm(\bo)=[  \boldsymbol{\mathcal{B}(\bo)}\pm\boldsymbol{\mathcal{B^*}(\bo)}]/2$.

The eigenvalues of $\mathcal{H}_B(\bo)$ are given by $\eta\epsilon(\bo)=[\epsilon_\mu(\bo), -\epsilon_\mu(\bo)]$, where \bea\epsilon_\mu(\bo)=\sqrt{A_\mu^2(\bo)-|B_\mu(\bo)|^2}\label{posi},\eea
$A_\mu$ and $B_\mu$ are the eigenvalues of $\boldsymbol{\mathcal{A}}(\bo)$ and $\boldsymbol{\mathcal{B}}(\bo)$ respectively. For the XY model $\mathcal{H}_B(\bo)$ is given by
\begin{figure}
\centering
  \subfigure[\label{XYband}]{\includegraphics[width=.45\linewidth]{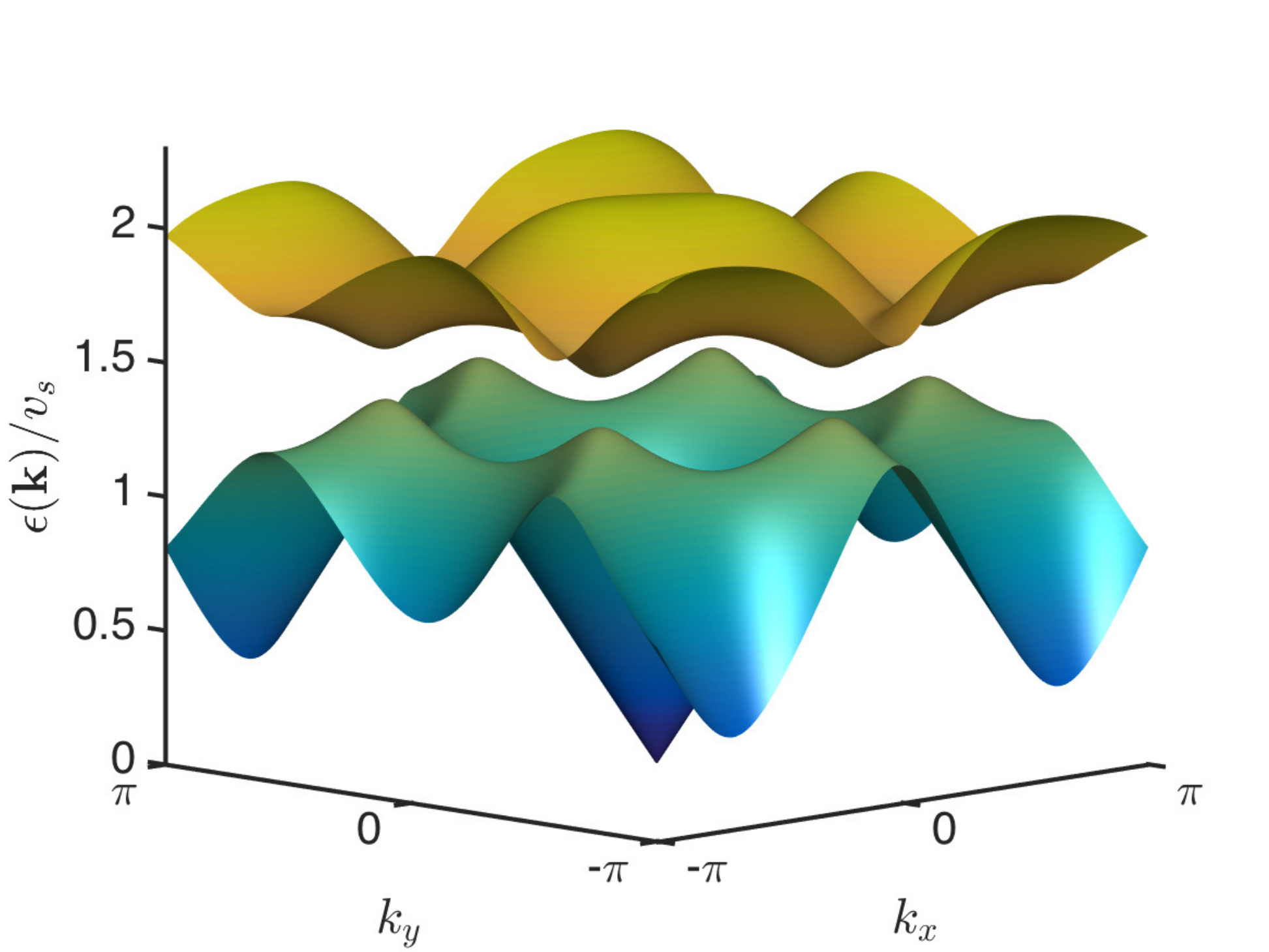}}
   \quad
   \subfigure[\label{HCB}]{\includegraphics[width=.45\linewidth]{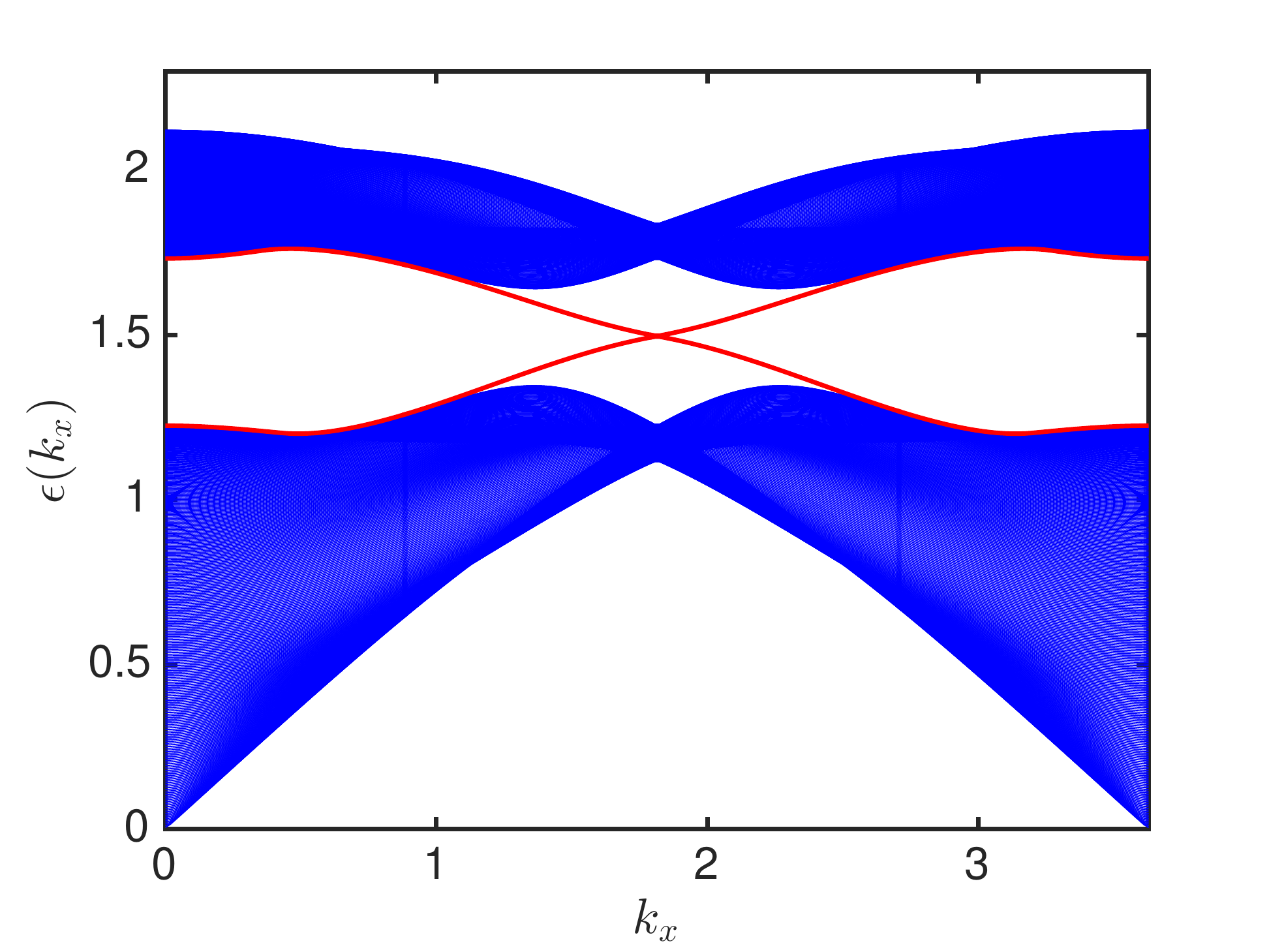}}
\caption{Color online. $(a)$ The band structure of the XY honeycomb ferromagnet.  $(b)$ The energy band for a one-dimensional strip on the honeycomb lattice. The parameters are    $J=0.5$,~$D=0.25J$, $S=1/2$. }
\end{figure}
 
 \begin{eqnarray}
\mathcal{H}_B(\bo)&= 3v_s[\sigma_z \boldsymbol{\mathcal{A}}(\bo)+i\sigma_y \boldsymbol{\mathcal{B}}(\bo)],
\label{ham3}
\end{eqnarray}
with $\boldsymbol{\mathcal{A}}(\bo)=\tau_0-\boldsymbol{\mathcal{B}}(\bo)$ and $\boldsymbol{\mathcal{B}}(\bo)=(\tau_+\gamma_\bo +h.c.)/2$.
The positive eigenvalues [Eq.~\ref{posi}] are given by
\begin{eqnarray}
\epsilon_\pm({\bo})=  3{v}_s\sqrt{ 1 \pm |\gamma_\bo|}.
\label{hg}
\end{eqnarray}
 The magnon excitations exhibit Dirac nodes at $\bold K_\pm$ with an energy of $3v_s$. To generate a gap,  we follow the same approach above. For simplicity, we ignore an external magnetic field and a NNN isotropic interaction $J^\prime$. Hence, the DM interaction that  would open a gap has to be parallel to the $x$-quantization axis,
 \begin{eqnarray}
 H_{DM}=D\sum_{\langle\langle lm\rangle\rangle}\nu_{lm}(S_l^yS_m^z -S_l^zS_m^y).
 \label{ham4}
 \end{eqnarray}
 
In the HP bosonic mapping, this corresponds to a magnetic flux of $\phi=\pi/2$. 
 In the $S_x$ quantization axis, $S_y$ and $S_z$ are off-diagonals. The momentum space Hamiltonian is given by
\begin{eqnarray}
\mathcal{H}^{DM}_B(\bo)= -v_D\sigma_0\tau_z\rho_\bo.
   \label{ham5}
\end{eqnarray}
The positive eigenvalues of the full Hamiltonian  are given by
\begin{eqnarray}
&\epsilon_\pm({\bo})=   \Bigg[ \lb 3{v}_s \pm\sqrt{(v_D\rho_\bo)^2 +\lb\frac{3v_s|\gamma_\bo|}{2}\rb^2}\rb^2-\lb\frac{3v_s|\gamma_\bo|}{2}\rb^2\Bigg]^{1/2},
\label{ham6}
\end{eqnarray}
At $\bold K_\pm$,  a gap  of magnitude $|\Delta|= 2|m|$ is generated as shown in Fig.~\ref{XYband}. Similar to the Heisenberg model, there exist magnon edge states in the vicinity of the bulk gap as depicted in Fig.~\ref{HCB}.
  
Surprisingly, Eqs.~\ref{ham0} and \ref{ham4} actually map  to interacting hardcore bosons on the honeycomb lattice via the Matsubara-Matsuda transformation \cite{mats} $S_l^x\to(b_l^\dg+b_l)/2;~S_l^y\to(b_l^\dg-b_l)/2i;~S_l^z=n_l-1/2$, where $n_l=b_l^\dg b_l$. The hardcore boson Hamiltonian is given by
\begin{eqnarray}
&H= -t\sum_{\langle lm\rangle}\lb b^\dg_l b_m +  h.c.\rb - t^\prime e^{-i\phi}\sum_{\langle\langle lm\rangle\rangle}\nu_{lm}\bigg[ (b_l^\dg-b_l) f_m-  (b_m^\dg-b_m) f_l\bigg],
 \label{ham6}
\end{eqnarray}
where $t\to J,~t^\prime\to D$, $f_l=n_l-1/2$, and $\phi=\pi/2$.   This model [Eq.~\ref{ham6}] offers a physical realization of these magnon edge states using ultracold atoms trapped in honeycomb optical lattices. Unfortunately, this model cannot be simulated by quantum Monte Carlo (QMC) methods due to a sign problem. However, it is amenable to other numerical simulations such as exact diagonalization. Introducing a magnetic field introduces additional phases into the system. For instance, a magnetic field along the $z$-axis introduces superfluid and  Mott insulating phases, whereas a staggered magnetic  field introduces a charge-density wave insulator.  In this case, the corresponding hard-core boson model can be written as 
\begin{eqnarray}
H&= -t\sum_{\langle lm\rangle}( b^\dg_l b_m +  h.c.) - t^\prime \sum_{\langle\langle lm\rangle\rangle}(e^{i\nu_{lm}\phi}b_l^\dg b_m +h.c.) -\sum_l (\mu +U_l) n_l,
\label{hamm6}
\end{eqnarray}
where $\mu$ is the chemical potential and $U_l$ is a staggered onsite potential on sublattice $A$ and $B$. They correspond to a (staggered) magnetic field along the $z$-axis in the spin language. For $t^\prime=0$,  Eq.~\ref{hamm6} is amenable to QMC simulation as recently shown \cite{alex9}.  Also recently, we have complemented  the  QMC results using the method presented in this paper \cite{sol}. Hence, the  HP spin wave method offers a simple approach to capture the topological properties  of bosonic models that cannot be simulated by QMC.

\subsection{Weyl magnon}

Finally, we address the Weyl magnons in 3D lattices. Recently, Weyl points were observed  on  breathing pyrochlore lattice governed by \cite{fei, up}   
\begin{eqnarray} H &=& J \sum_{\langle{ij}\rangle
  \in \rm{u}} {\bf S}_i \cdot {\bf S}_j
+ J' \sum_{\langle{ij}\rangle \in \rm{d}} {\bf S}_i \cdot {\bf S}_j  + K \sum_{i} \lb {\bf S}_i \cdot \hat{z}_i \rb^2, 
\label{eq1} 
\end{eqnarray}
where $J>0$ and $J'>0$ are the exchange couplings between the nearest-neighbour
spins on the up-pointing and down-pointing tetrahedra respectively (see  Ref. \cite{fei}), and  $D$ is a single-ion anisotropy. It can be easy-axis ($K<0$) or easy-plane ($K>0$). In the former case, the spins would prefer the $z$-axis; whereas in the latter case the $xy$ plane is  auspicious. 

Although a comprehensive analysis of this model has been studied in Ref. \cite{fei}, the criteria for the existence of Weyl magnons were  not mentioned  and understood properly. Here, we argue that breaking of pseudo spin $\mathcal{T}$-symmetry is a condition for Weyl points to exist in quantum magnetic systems.  The  linear spin wave theory Hamiltonian  derived in  Ref. \cite{fei}  has the general form given in Eq.~\ref{main}. Now, the Bogoliubov Hamiltonian can be cast into  the form of Eq.~\ref{Bogoliubovb}. From this equation, we see that the momentum space Hamiltonian resembles that of electronic systems. 

In addition to Weyl nodes obtained along the BZ paths for $K>0$ and $J\neq J'$ \cite{fei}, there is additional non-degenerate band-touching points at the corners of the BZ. The system should realize Dirac Hamiltonian at the corners of the BZ as shown above and also a Weyl Hamiltonian near the Weyl points.   To check whether pseudo spin ${\mathcal T}$-symmetry  is preserved or broken at the Dirac  or Weyl points respectively, one should follow the approach outlined above. Basically, one has to expand $\boldsymbol{\mathcal{A}}(\bo)$ and $\boldsymbol{\mathcal{B}}_\pm(\bo)$ near the band-touching points and project the resulting Hamiltonian onto the bands.   In principle, the Bogoliubov Hamiltonian [Eq.~\ref{Bogoliubovb}]  near the Weyl points should break ${\mathcal T}$-symmetry. Therefore, one recovers the usual criteria for  Weyl semimetals \cite{aab1}. In contrast to 2D systems, a gap is not needed to observe  edge states in 3D Weyl magnons. Edge states exist in 3D Weyl magnons provided the momentum lies between the Weyl nodes \cite{fei}.
\section{Conclusion}
In summary, we have shown that physical realistic models of honeycomb quantum spin magnets exhibit nontrivial topology in the magnon excitations. In 2D ordered honeycomb  quantum magnets, we showed that the non-degenerate band-touching points (at the corners of the Brillouin zone) in  the magnon excitation spectrum  realize a massless Dirac Hamiltonian. Opening of a gap  at the Dirac points requires the breaking of inversion symmetry of the lattice. This leads to nontrivial topological magnon insulator with magnon edge states  propagating on the edges of the material, similar to topological insulators in electronic systems.  These magnon edge states also manifest in hardcore bosons on honeycomb lattice.  The hardcore boson model proposed in Eqs.~\ref{ham6} and \ref{hamm6} should be studied by numerical approach to further substantiate the existence of magnon edge modes in this system.  Since there are many physical 2D honeycomb quantum magnetic materials in nature, these  results suggest new experiments in ordered quantum magnets and ultracold atoms in honeycomb  optical lattices, to search for magnon Dirac materials and topological magnon insulators on the honeycomb lattice.   For 3D ordered quantum magnets, Weyl points are possible in the magnon excitations \cite{fei}. We argued that the Bogoliubov Hamiltonian near the Weyl points should yield a low-energy Hamiltonian that breaks time-reversal symmetry of the pseudo spins.  At nonzero temperature and external magnetic field, there is a possibility of topological magnon Hall effect \cite{hka, xc} and spin Nernst effect \cite{hka1}, similar to the kagome, Lieb, and  pyrochlore lattices \cite{hka, xc, hka1}. The analysis of magnon Hall effect for the honeycomb lattice will be reported elsewhere.  

\section{Acknowledgments}  The author would like to thank African Institute for Mathematical  Sciences. Research at Perimeter Institute is supported by the Government of Canada through Industry Canada and by the Province of Ontario through the Ministry of Research
and Innovation.


\begin{thebibliography}{10}
\bibitem{up}
Upon finishing this work we became aware of Ref.~\cite{fei}, which focuses on  Weyl points on the breathing pyrochlore lattice.
\end{thebibliography}
\end{document}